\documentstyle[12pt,psfig]{article}
\let\section=\subsection     \let\subsection=\subsubsection                

\makeatletter
\newbox\slashbox \setbox\slashbox=\hbox{\large$/$}
\def\pslash#1{\setbox\@tempboxa=\hbox{$#1$}
  \@tempdima=0.5\wd\slashbox \advance\@tempdima 0.5\wd\@tempboxa
  \copy\slashbox \kern-\@tempdima \box\@tempboxa}
\def\slash{\protect\pslash}
\def\openone{\leavevmode\hbox{\small1\kern-3.3pt\normalsize1}}
\makeatother

\begin{document}
\begin{center}
  {\large \bf THE CHIRAL PHASE TRANSITION, RANDOM}\\[2mm]
  {\large \bf MATRIX MODELS, AND LATTICE DATA}\\[5mm]
  T.~WETTIG$^{1,2}$, T.~GUHR$^1$, A.~SCH\"AFER$^3$, and
  H.~A.~WEIDENM\"ULLER$^1$\\[5mm] 
  {\small \it $^1$Max-Planck-Institut f\"ur Kernphysik, Postfach
    103980,\\ D-69029 Heidelberg, Germany\\
    $^2$Institut f\"ur Theoretische Physik, Technische Universit\"at
    M\"unchen,\\ D-85747 Garching, Germany\\
    $^3$Institut f\"ur Theoretische Physik, Universit\"at Regensburg,\\
    D-93040 Regensburg, Germany}\\[8mm]
\end{center}

\begin{abstract}\noindent
  We present two pieces of evidence in support of the conjecture that
  the microscopic spectral density of the Dirac operator is a
  universal quantity.  First, we compare lattice data to predictions
  from random matrix theory.  Second, we show that the functional form
  of the microscopic spectral correlations remains unchanged in random
  matrix models which take account of finite temperature.
  Furthermore, we present a random matrix model for the chiral
  phase transition in which all Matsubara frequencies are included.
\end{abstract}

\section{Introduction}
\label{sec1}

Motivated by the Leutwyler-Smilga sum rules for the inverse powers of
the eigenvalues of the Dirac operator in a finite volume
\cite{Leut92}, Shuryak and Verbaarschot \cite{Shur93} introduced the
so-called microscopic spectral $\rho_s$ density of the Dirac operator
$i\slash{D}$.  This quantity is defined as
\begin{equation}
  \label{eq1.1}
  \rho_s(z)=\lim_{V\rightarrow\infty}\frac{1}{V\Sigma}
  \rho(\frac{z}{V\Sigma}) \: ,
\end{equation}
where $\rho(\lambda)=\sum_n\delta(\lambda-\lambda_n)$ is the spectral
density of the Dirac operator, $V$ is the space-time volume, and
$\Sigma$ is the chiral condensate.  The microscopic spectral density
carries information on how the thermodynamic limit is approached
which, among other things, is very useful for the analysis of lattice
data, in particular in the chiral limit.  It was conjectured that
$\rho_s$ should be a universal quantity, i.e., that it is insensitive
to the details of the dynamics.  As a universal quantity, $\rho_s$
should depend only on the symmetries of the problem and, therefore, be
calculable in a much simpler theory than QCD, namely random matrix
theory (RMT).  In RMT, the matrix of the Dirac operator (including a
mass term) in a chiral basis in Euclidean space is replaced by a
random matrix according to
\begin{equation}
  \label{eq1.2}
  \left[\matrix{im & i\slash{D}\cr (i\slash{D})^\dagger&im}\right]
  \quad\longrightarrow\quad 
  \left[\matrix{im & W\cr W^\dagger&im}\right] \:,
\end{equation}
where $W$ is a random matrix of dimension $N$.  The average over gauge
field configurations is replaced by the average over random matrices,
and the weighting function $\exp(-S_{\rm glue})$ is replaced by a
simple Gaussian distribution $P(W)\sim\exp(-N\Sigma^2{\,\rm Tr\,}
WW^\dagger)$ of the random matrices.  Depending on the representation
of the fermions and the number of colors, there are three different
universality classes which have been classified in
Ref.~\cite{Verb94a}.  They are described by the three chiral Gaussian
ensembles: the chiral Gaussian Orthogonal Ensemble (chGOE) where $W$
is real, the chiral Gaussian Unitary Ensemble (chGUE) where $W$ is
complex, and the chiral Gaussian Symplectic Ensemble (chGSE) where $W$
is quaternion real.  The microscopic spectral density has been
computed for all three ensembles in the framework of RMT
\cite{Verb93,Naga95}.

There were a number of findings which added support to the
universality conjecture for $\rho_s$.  We refer the reader to a recent
review by Verbaarschot \cite{Verb96b} which also gives a comprehensive
account of other applications of RMT in QCD.  In this paper, we
present direct evidence for the universality of $\rho_s$ by comparing
lattice data to RMT predictions.  This is done in Sec.~\ref{sec2.1}.
In Sec.~\ref{sec2.2} we add another piece of evidence by showing that
the functional form of $\rho_s$ and of the microscopic spectral
correlations in general does not change in random matrix models which
take account of finite temperature.  In Sec.~\ref{sec3} we discuss a
model for the finite-temperature chiral phase transition in which
all Matsubara frequencies are included.  Conclusions are drawn in
Sec.~\ref{sec4}.

\section{Universality of $\rho_s$ and of the microscopic spectral
  correlations} 
\label{sec2}

\subsection{Lattice data and predictions from random matrix theory}
\label{sec2.1}

We analyze lattice data obtained by Berbenni-Bitsch and Meyer for an
SU(2) gauge theory with staggered fermions in the quenched
approximation with $\beta=2.0$.  To study finite-volume effects, data
were obtained for four different lattice sizes: 4$^4$, 6$^4$, 8$^4$,
and 10$^4$.  For each lattice, the complete spectrum of the Dirac
operator was obtained for a very large number of configurations (9979,
9953, 3896, and 477 configurations for the 4$^4$, 6$^4$, 8$^4$, and
10$^4$ lattice, respectively).  This was necessary to obtain good
statistics.  The numerical eigenvalues of the Dirac operator satisfy
an analytical sum rule with excellent precision (relative accuracy
$\sim 10^{-8}$).

According to Ref.~\cite{Verb94a}, staggered fermions in SU(2) should
be described by the chGSE.  Analytical results exist for the
distribution of the smallest eigenvalue (only for the quenched
approximation) \cite{Forr93},
\begin{equation}
  \label{eq2.1.1}
  P(\lambda_{\rm min})=\sqrt{\frac{\pi}{2}}c(c\lambda_{\rm min})^{3/2}
  I_{3/2}(c\lambda_{\rm min})e^{-\frac{1}{2} (c\lambda_{\rm min})^2} \:,
\end{equation}
and for the microscopic spectral density (also for the unquenched
case) \cite{Naga95},
\begin{equation}
  \label{eq2.1.2}
  \rho_s(z)=2\pi(cz)^2\int\limits_0^1 du \, u^2\int\limits_0^1 dv\, 
  \left[J_{a-1}(2uvcz)J_a(2ucz)-vJ_{a-1}(2ucz)J_a(2uvcz)\right] \:.
\end{equation}
Hence, a direct comparison with lattice data is possible.  Here,
$a=N_f+2\nu+1$, where $N_f$ is the number of massless dynamical
quarks and $\nu$ is the topological charge.  The parameter $c$ sets
the scale on which the eigenvalue is measured.  It should be pointed
out that $c$ is not a free parameter but determined by the lattice
volume and the chiral condensate according to Eq.~(\ref{eq1.1}).

In Figure~1 we have plotted the data and the predictions from RMT for
both $P(\lambda_{\rm min})$ and $\rho_s(z)$ for all available lattice
sizes except for the 10$^4$ lattice where the number of configurations
was too small to obtain sufficiently good statistics.  The parameter
$a$ in (\ref{eq2.1.2}) is 1 since $N_f=0$ and since $\nu_{\rm eff}=0$
on the lattice \cite{Verb96a}.  The data agree almost perfectly with
the RMT predictions for all but the smallest lattice.  The range over
which $\rho_s$ agrees with the RMT prediction increases with the
lattice size which is evident by the definition of $\rho_s$ in
Eq.~(\ref{eq1.1}).  We conclude that a lattice size of $6^4$ is
sufficient to identify the universal microscopic spectral density of
the Dirac operator.

\begin{figure}
\vspace*{-5mm}
\begin{center}
\begin{tabular}{cc}
\psfig{figure=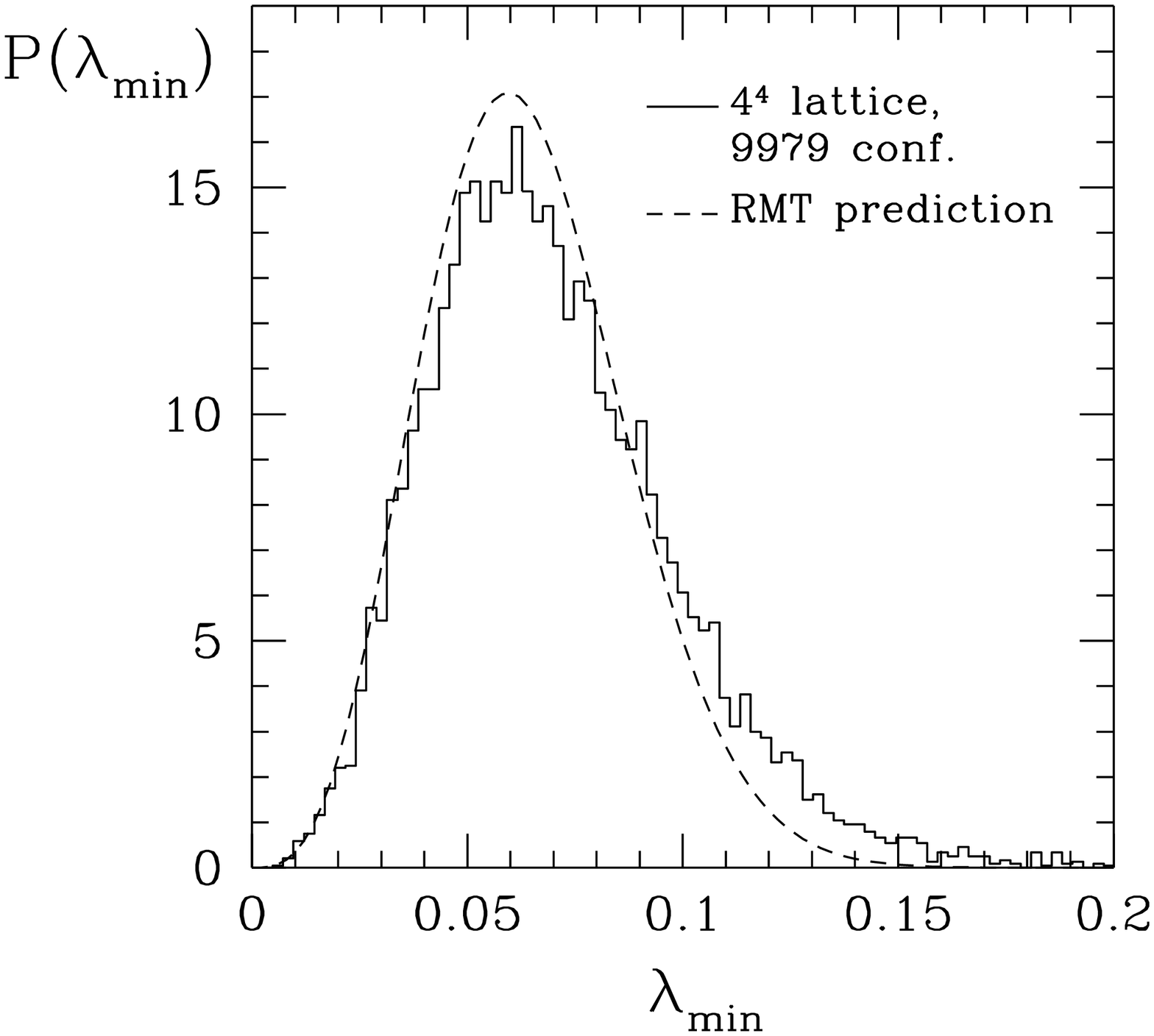,width=64mm}&
\psfig{figure=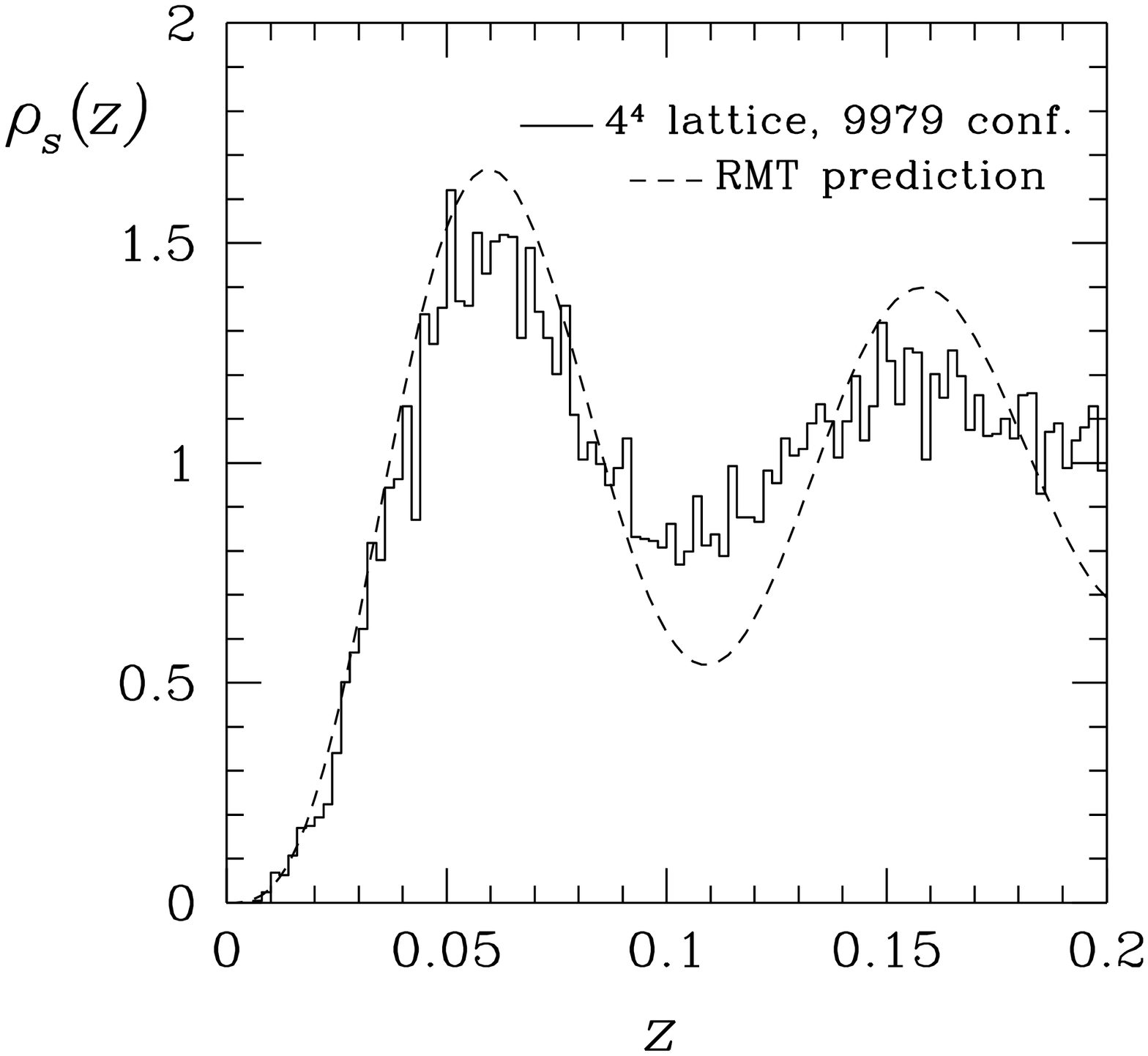,width=64mm}\\[-5mm]
\psfig{figure=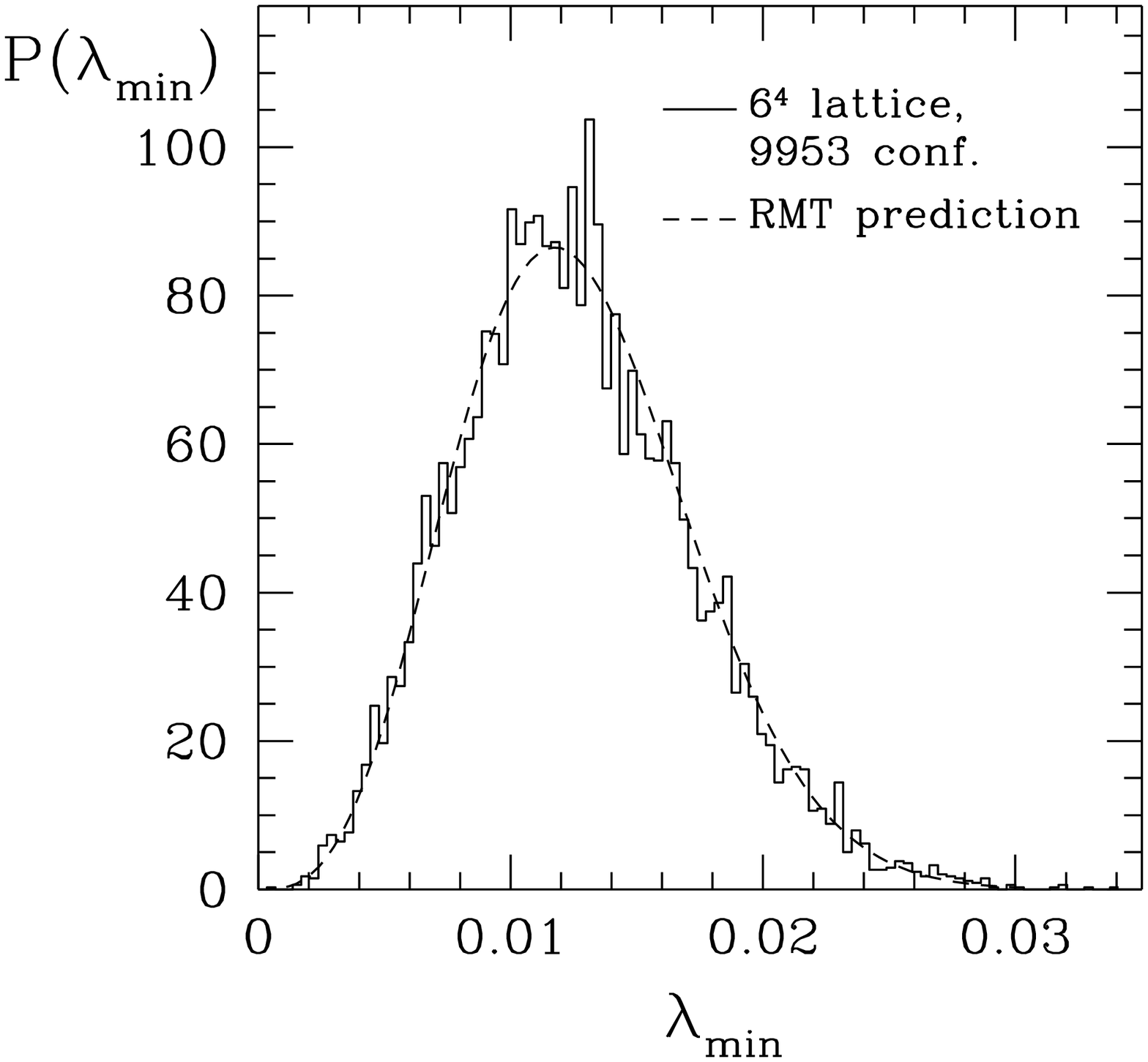,width=64mm}&
\psfig{figure=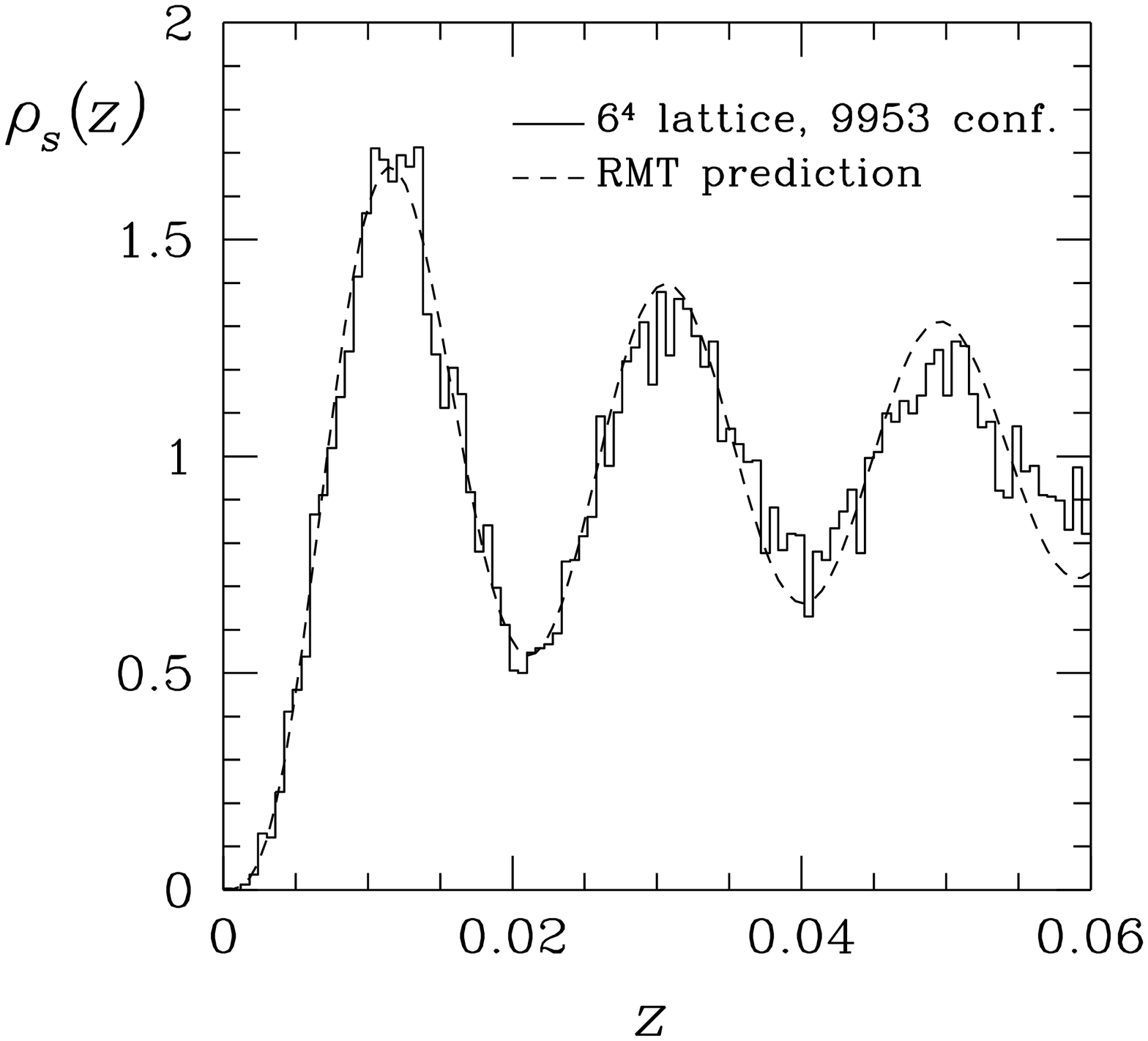,width=64mm}\\[-5mm]
\psfig{figure=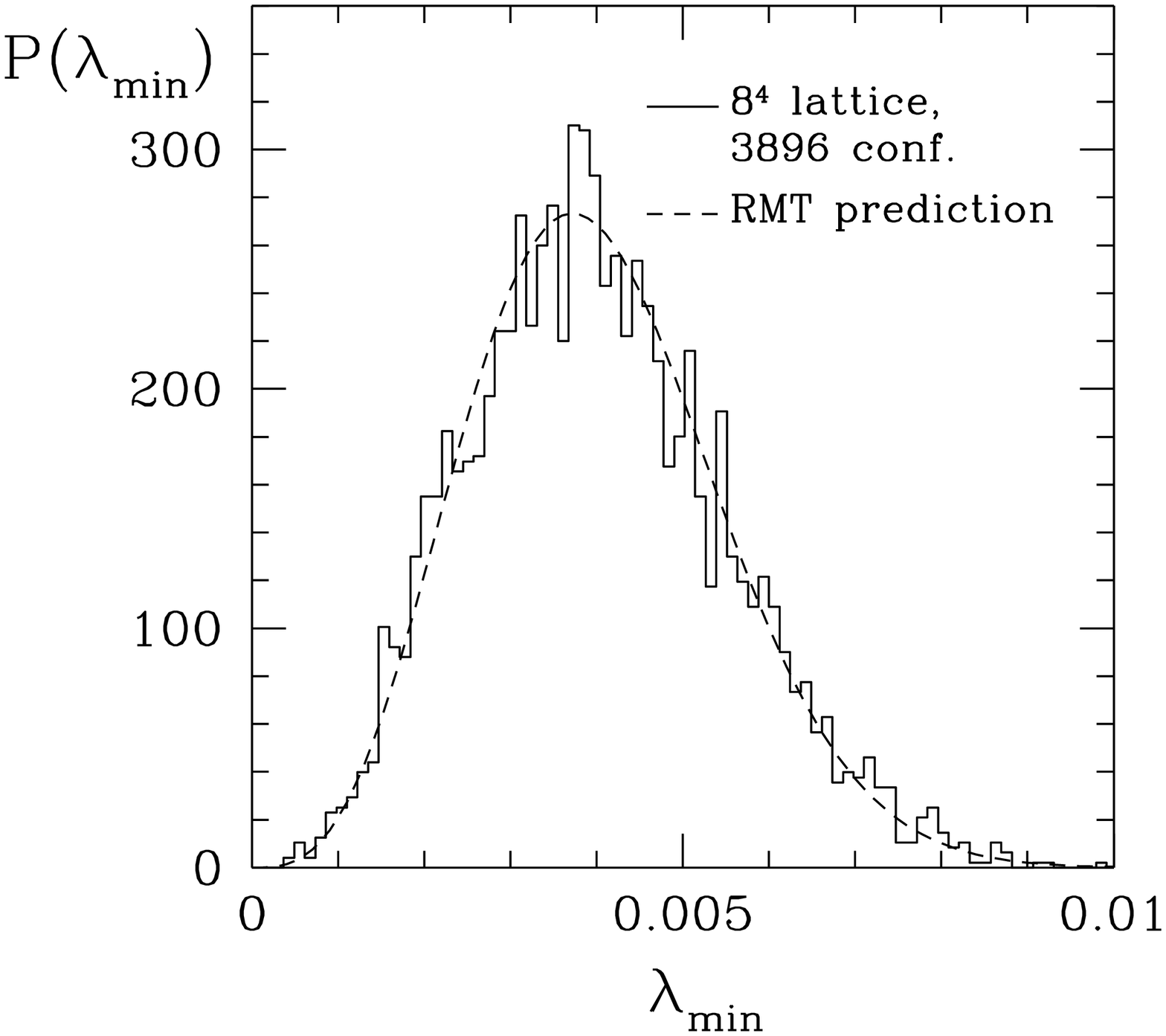,width=64mm}&
\psfig{figure=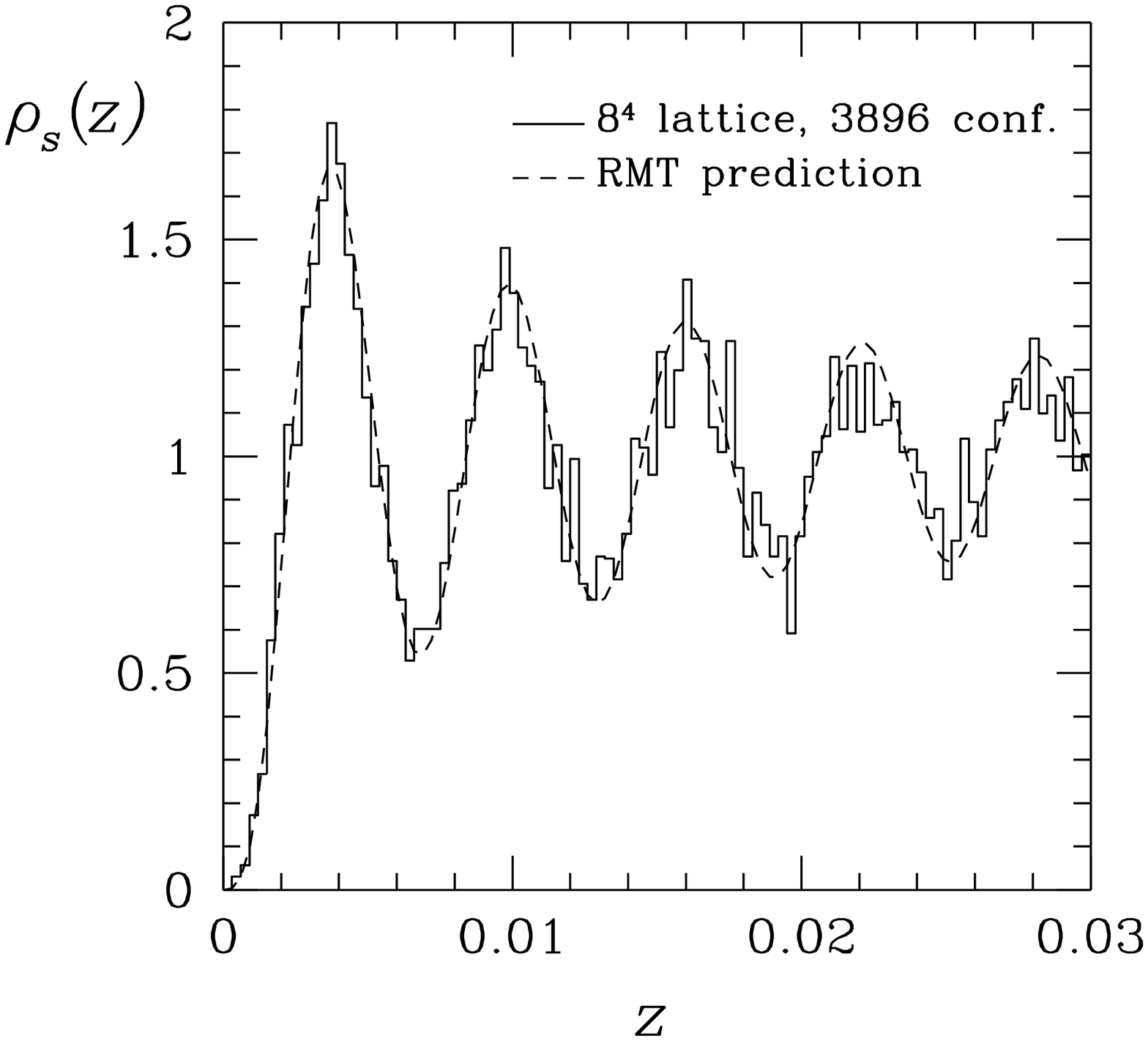,width=64mm}
\end{tabular}
\end{center}
\vspace*{-10mm}
\caption{Distribution of the smallest eigenvalue $P(\lambda_{\rm
    min})$ and microscopic spectral density $\rho_s(z)$ for three
  different lattice sizes (see text).  Note that there are no free
  parameters in the RMT prediction.}
\end{figure}

Unquenched calculations and their analysis in terms of RMT are in
pro\-gress.  It will be very interesting to see the effect of a small
dynamical quark mass in $\rho_s$.

\subsection{The microscopic spectral density at finite temperature}
\label{sec2.2}

Various random matrix models have been constructed for the spectrum of
the Dirac operator at finite temperature to describe generic features
of the chiral phase transition.  We shall discuss these models in more
detail in Sec.~\ref{sec3}.  The common feature of all these models is
that a temperature-dependent matrix $Y$ is added to the random matrix
$W$ in Eq.~(\ref{eq1.2}).  The specific form of $Y$ depends on the
choice of basis.  However, as we shall see in Sec.~\ref{sec3}, the
matrix $Y$ can always be chosen real and diagonal.  This defines the
generic problem.

The matrix of the Dirac operator (for $m=0$) thus has the form
\begin{equation}
  \label{eq2.2.1}
  \left[\matrix{0&W+Y\cr W^\dagger+Y&0}\right]
\end{equation}
with $Y={\rm diag}(Y_1,\ldots,Y_N)$.  The $Y_n$ are real and depend on
$T$.  The dimension $N$ of the matrix $W$ can be identified with the
space-time volume $V$.  For the moment, let us concentrate on the
chGUE and on the quenched approximation.  In this case, we can compute
the chiral condensate which is given by
\begin{equation}
  \label{eq2.2.2}
  \Xi(T)=\Sigma^2\bar x \: ,
\end{equation}
where $\Sigma$ is the chiral condensate at $T=0$ and $\bar x$ is the
only real and positive solution of
\begin{equation}
  \label{eq2.2.3}
  \Sigma^2=\frac{1}{N}\sum_{n=1}^N\frac{1}{{\bar x}^2+Y_n^2}
\end{equation}
or zero if no such solution exists \cite{Wett96a}.  The computation of
the microscopic spectral density and the corresponding microscopic
spectral correlations is more difficult since the standard methods
involving orthogonal polynomials cannot be applied here due to the
presence of the additional matrix $Y$.  We have used the graded
eigenvalue method \cite{Guhr91} and extended it to the present case
with the additional chiral symmetry of the Dirac operator
\cite{Guhr96,Guhr97}.  A similar calculation has been done
simultaneously and independently by the authors of Ref.~\cite{Sene97}.

The result of the calculation is very simple: The functional form of
$\rho_s$ and of the microscopic spectral correlations remains
unchanged provided that we take into account the temperature
dependence of the chiral condensate \cite{Jack96b}.  Specifically, we
find that the $k$-point functions can be written as
\begin{equation}
  \label{eq2.2.4}
  R_k(x_1,\ldots,x_k)=\det[C(x_p,x_q)]_{p,q=1,\ldots,k} \: ,
\end{equation}
where in the microscopic limit the function $C$ is given by
\begin{equation}
  \label{eq2.2.5}
  C(x_p,x_q)=2N\Xi u_p\frac{u_pJ_1(u_p)J_0(u_q)-u_qJ_0(u_p)J_1(u_q)}
    {u_p^2-u_q^2} 
\end{equation}
with $u_p=2N\Xi x_p$ and $u_q=2N\Xi x_q$.  Hence, the only change with
respect to the zero-temperature case is that the microscopic variables
are rescaled by $\Xi$, the temperature-dependent generalization of
$\Sigma$.  This result adds further support to the conjecture that
$\rho_s$ (and the microscopic spectral correlations) are universal.

So far, the calculation has been done for the quenched approximation
only.  The unquenched case gives rise to some technical complications,
but we have no doubt that the same result will be obtained: The
functional form of the microscopic spectral density and the
microscopic spectral correlations will not change provided that the
arguments are rescaled by the temperature-dependent chiral condensate.
We make the same conjecture for the chGOE and the chGSE although it is
not clear at the moment how the calculation should be done in these
two cases since the graded eigenvalue method requires the computation
of an Itzykson-Zuber-like integral which is not yet known for the
chGOE and the chGSE.  Since the microscopic spectral correlations are
known at zero temperature for all three chiral ensembles (even in the
presence of massless dynamical quarks) and since the temperature
dependence of the chiral condensate can be computed easily for each
ensemble, the finite-temperature results thus follow immediately.

\section{A random matrix model for the chiral phase transition with
  all Matsubara frequencies included}
\label{sec3}

As we have already mentioned in the previous section, at finite
temperature a temperature-dependent matrix $Y$ has to be added to the
random matrix $W$ representing the matrix elements of the Dirac
operator.  Why this is so will become obvious below.  Since $P(W)$ is
invariant under a transformation of the basis, we can always make $Y$
diagonal by a suitable basis transformation.  While at zero
temperature the choice of basis was completely irrelevant we now have
to specify the basis with respect to which the matrix elements of the
Dirac operator are evaluated.  A different choice of basis will be
reflected in a different form of the matrix $Y$.  In
Refs.~\cite{Jack95,Step96a} a plane wave basis was chosen whereas the
basis of Ref.~\cite{Wett96a} consisted of (anti-) instanton zero
modes.  In \cite{Jack95}, only the lowest Matsubara frequency was
included in $Y$ since it is sufficient to determine the properties of
the chiral phase transition at $T=T_c$.  We now present a model which
includes all Matsubara frequencies, thereby extending the validity of
the model to lower temperatures.\footnote{After completion of this
  work, we were informed by M.~A.~Nowak that an equivalent model,
  derived from the NJL model and additionally including a chemical
  potential, was discussed recently in Ref.~\cite{Jani97}.  One
  can show that the two models yield identical results.} (For a
comparison with the model presented in Ref.~\cite{Step96a} see below.)

If one chooses a plane-wave basis $\phi_k({\bf x})e^{i\omega_k\tau}$
at finite temperature, the $\partial_0$-term in $i\slash{D}$ gives
rise to a term proportional to $\omega_k=(2k+1)\pi T$ ($k$ integer) in
the matrix of the Dirac operator.  These terms are included in $Y$.
The basis states $\phi_k({\bf x})$ corresponding to the same Matsubara
frequency $\omega_k$ are coupled by a random matrix.  In the same way,
basis states belonging to different Matsubara frequency should be
coupled by random matrices.  The matrix of the Dirac operator at
finite temperature can thus be written in the form
\begin{equation}
  \label{eq3.1}
  \left[\matrix{im&W+\Omega\cr W^\dagger+\Omega&im}\right]
\end{equation}
with
\begin{equation}
  \label{eq3.2}
  \Omega={\rm diag}(\pi T,\ldots,\pi T,-\pi T,\ldots,-\pi T,
  3\pi T,\ldots,3\pi T,-3\pi T,\ldots,-3\pi T,\ldots) \:.
\end{equation}
The dimension of a submatrix of $\Omega$ belonging to one single
Matsubara frequency is equal to $K$ which should be identified with
the three-volume $V_3$.  The four-volume is $V_4=V_3/T=K/T$.  The
elements of the random matrix $W$ are distributed according to
$P(W)\sim\exp(-V_4\Sigma^2{\,\rm Tr\,}WW^\dagger)$.  The model
presented in Ref.~\cite{Step96a} also took into account all Matsubara
frequencies.  However, basis states corresponding to different
Matsubara frequencies were coupled by the same random matrix.  There
is no reason why this should be the case, and the resulting
temperature dependence of the chiral condensate shows non-analytic
features which are physically unrealistic.

Going through the same steps as in \cite{Wett96a}, we obtain a
saddle-point equation similar to (\ref{eq2.2.3}) from which the chiral
condensate follows as $\langle\bar qq\rangle=\Sigma^2\bar x$.  This
equation reads (in the chiral limit $m=0$)
\begin{equation}
  \label{eq3.3}
  \Sigma^2=\frac{1}{V_4}\sum_{n=1}^N\frac{1}{{\bar x}^2+\Omega_n^2}
  =\frac{2K}{K/T}\sum_{k=0}^{N/2K}\frac{1}{{\bar x}^2+\omega_k^2} \:,
\end{equation}
where the upper limit $N/2K$ of the sum should go to infinity.  We now
perform the sum, choose units in which $\langle\bar
qq\rangle(T\!=\!0)=1$, and express the result in terms of $\langle\bar
qq\rangle$ to obtain
\begin{equation}
  \label{eq3.4}
  \langle\bar qq\rangle=\tanh\frac{\langle\bar qq\rangle}{T/T_c} \:.
\end{equation}
This equation can only be solved numerically, and the result is
plotted in Figure~2.  The low-temperature behavior of the chiral
condensate can be obtained analytically in this model.  We have
\begin{equation}
  \label{eq3.5}
  \langle\bar qq\rangle\approx 1-2\exp\left(-\frac{2}{T/T_c}\right) 
  \qquad {\rm as} \;\; T\rightarrow 0 \:.
\end{equation}
We observe that the chiral condensate does not change significantly
unless the temperature is close to $T_c$.  While this is in
qualitative agreement with lattice data \cite{Kars95}
Eq.~(\ref{eq3.5}) does not agree with the well-established
low-temperature expansion of chiral perturbation theory \cite{Gass87}.
The reason for this discrepancy is that the random-matrix model in its
present form is effectively zero-dimensional in space and, therefore,
unable to describe the effect of the Goldstone bosons adequately.  In
order to account for the Goldstone bosons, additional non-random parts
would have to be added to the random matrix.

\begin{figure}[t]
\vspace*{-10mm}
\centerline{\psfig{figure=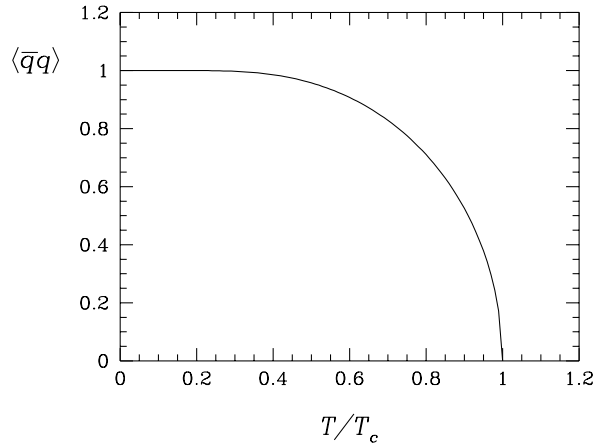,width=80mm}}
\caption{The chiral condensate in units of its value at $T=0$ for the
model discussed in the text.}
\end{figure}

At the critical temperature, we can compute the critical exponents
$\beta$, $\gamma$, and $\delta$.  These are defined using the reduced
temperature $t=(T-T_c)/T_c$ by $\langle\bar qq\rangle\sim|t|^\beta$,
$\chi\sim |t|^{-\gamma}$ with $\chi=\partial\langle\bar qq\rangle/
\partial m|_{m=0}$, and $\langle\bar qq\rangle\sim m^{1/\delta}$ at
$T=T_c$.  We obtain $\beta=1/2$, $\gamma=1$, and $\delta=3$ in
agreement with \cite{Jack95}.  This supports the position taken in
\cite{Jack95} that only the lowest Matsubara frequency is responsible
for the critical behavior at the phase transition.

\section{Conclusions}
\label{sec4}

We have presented very direct evidence for the universality of the
microscopic spectral density of the Dirac operator by comparing
lattice data obtained by Berbenni-Bitsch and Meyer to predictions from
random matrix theory.  Impressive agreement was found already for
relatively small lattices.  Clearly, in the microscopic limit the
lattice data are described by random matrix theory.  Thus, RMT is a
very useful tool in the analysis of lattice data and may provide
guidance for further lattice calculations.  It remains to be seen
whether RMT can also prove useful with respect to simplifying the
generation of gauge field configurations.  This would be enormously
helpful, in particular close to the chiral limit.  

The universality of the microscopic spectral correlations of the Dirac
operator was further supported by a generic finite-temperature
calculation in which it was found that the functional form of the
microscopic spectral correlations does not change with temperature
provided that the temperature dependence of the chiral condensate is
taken into account in the microscopic scale.  Although this
calculation was only done for the chGUE in the quenched approximation,
it is more than likely that analogous results will be obtained in the
unquenched case and for the other two chiral ensembles.

A model was presented for the chiral phase transition at finite
temperature.  This model took account of all Matsubara frequencies.
It reproduced the observation from lattice data that the chiral
condensate remains almost constant up to temperatures close to $T_c$
and also confirmed the position of Ref.~\cite{Jack95} that the lowest
Matsubara frequency is responsible for the properties of the chiral
phase transition.  Reasons for discrepancies with the low-temperature
expansion of chiral perturbation theory have been discussed.

A number of very interesting topics such as the effect of a chemical
potential \cite{Step96b,Jani96a} and the question of topological
charge and the axial U(1) anomaly \cite{Jani96b} have not even been
mentioned in this article.  Although the field of applying RMT to
problems of QCD is a very recent one, many useful applications have
already been found, and we are convinced that this will not cease to
be the case in future work.

\section*{Acknowledgments}

We thank M.~E.~Berbenni-Bitsch and S.~Meyer for providing us with
their lattice data and interesting discussions.  Jac Verbaarschot is
thanked for stimulating and helpful communication, and M.~A.~Nowak for
informing us about the model discussed in Ref.~\cite{Jani97}.


\begin{thebibliography}{99}
\itemsep=0cm

\bibitem{Leut92} H.~Leutwyler and A.~Smilga, Phys.~Rev.~{\bf D 46}
  (1992) 5607;

\bibitem{Shur93} E.~V.~Shuryak and J.~J.~M.~Verbaarschot,
  Nucl.~Phys.~{\bf A 560} (1993) 306;

\bibitem{Verb94a} J.~J.~M.~Verbaarschot, Phys.~Rev.~Lett.~{\bf 72} (1994)
  2531; 

\bibitem{Verb93} J.~J.~M.~Verbaarschot and I.~Zahed,
  Phys.~Rev.~Lett.~{\bf 70} (1993) 3852;
  J.~J.~M.~Verbaarschot, Nucl.~Phys.~{\bf B 426} (1994) 559;

\bibitem{Naga95} T.~Nagao and P.~J.~Forrester, Nucl.~Phys.~{\bf B 435}
  (1995) 401;

\bibitem{Verb96b} J.~J.~M.~Verbaarschot, hep-lat/9606009;

\bibitem{Forr93} P.~J.~Forrester, Nucl.~Phys.~{\bf B 402} (1993) 709;

\bibitem{Verb96a} J.~J.~M.~Verbaarschot, Phys.~Lett.~{\bf B 368} (1996)
  137;

\bibitem{Wett96a} T.~Wettig, A.~Sch\"afer, and H.~A.~Weidenm\"uller,
  Phys.~Lett.~{\bf B 367} (1996) 28;

\bibitem{Guhr91}  T.~Guhr, J.~Math.~Phys.~{\bf 32} (1991) 336;

\bibitem{Guhr96} T.~Guhr and T.~Wettig, J.~Math.~Phys.~{\bf 37} (1996)
  6395;

\bibitem{Guhr97} T.~Guhr and T.~Wettig, to be published;

\bibitem{Sene97} A.~D.~Jackson, M.~K.~\c Sener and
  J.~J.~M.~Verbaarschot, to be published;

\bibitem{Jack96b} A.~D.~Jackson, M.~K.~\c Sener, and
  J.~J.~M.~Verbaarschot, Nucl.~Phys.~{\bf B 479} (1996) 707;

\bibitem{Jack95} A.~D.~Jackson and J.~J.~M.~Verbaarschot,
  Phys.~Rev.~{\bf D 53} (1996) 7223;

\bibitem{Step96a} M.~A.~Stephanov, Phys.~Lett.~{\bf B 375} (1996) 249;

\bibitem{Jani97}R.~A.~Janik, M.~A.~Nowak, and I.~Zahed,
  Phys.~Lett.~{\bf B 392} (1997) 155; see also M.~A.~Nowak, M.~Rho and
  I.~Zahed, {\em Chiral Nuclear Dynamics}, World Scientific
  (Singapore) 1996;

\bibitem{Kars95} F.~Karsch, Nucl.~Phys.~{\bf A 590} (1995) 367;

\bibitem{Gass87} J.~Gasser and H.~Leutwyler, Phys.~Lett.~{\bf B 184}
  (1987) 83; Phys.~Lett.~{\bf B 188} (1987) 477;

\bibitem{Step96b} M.~A.~Stephanov, Phys.~Rev.~Lett.~{\bf 76} (1996)
  4472;

\bibitem{Jani96a} R.~A.~Janik, M.~A.~Nowak, G.~Papp, and I.~Zahed,
  Phys.~Rev.~Lett.~{\bf 77} (1996) 4876 and cond-mat/9612240;
  R.~A.~Janik, M.~A.~Nowak, G.~Papp, J.~Wambach, and I.~Zahed,
  hep-ph/9609491;

\bibitem{Jani96b} R.~A.~Janik, M.~A.~Nowak, G.~Papp, and I.~Zahed,
  hep-lat/9611012. 
\end{thebibliography}
\end{document}